\documentclass[journal]{IEEEtran} 
\IEEEoverridecommandlockouts
\usepackage{cite}
\usepackage{amsmath,amssymb,amsfonts}
\usepackage{algorithmic}
\usepackage[ruled,vlined]{algorithm2e}
\usepackage{diagbox}
\usepackage{graphicx}
\usepackage{textcomp}
\def\BibTeX{{\rm B\kern-.05em{\sc i\kern-.025em b}\kern-.08em
    T\kern-.1667em\lower.7ex\hbox{E}\kern-.125emX}}
\begin{document}

\title{FeDiSa: A Semi-asynchronous Federated Learning Framework for Power System Fault and Cyberattack Discrimination\\

\thanks{This work is the result of the research project funded by the Faculty of Science, Engineering and Built Environment (SEBE), Deakin University under the scheme 'Mini ARC Analogue Program (MAAP)'.}
}
\author{
    \IEEEauthorblockN{M. A. Husnoo$^\dag$}
, \IEEEauthorblockN{A. Anwar$^\dag$}%
    , \IEEEauthorblockN{H. T. Reda$^\dag$}, 
    \IEEEauthorblockN{N. Hosseinzadeh$^\ddagger$}
    , \IEEEauthorblockN{S. N. Islam$^\ddagger$}
    , \IEEEauthorblockN{A. N. Mahmood$^\ast$}
    and \IEEEauthorblockN{R. Doss$^\dag$}
    \\
    \IEEEauthorblockA{\textit{$^\dag$Centre for Cyber Security Research and Innovation, School of IT, Deakin University, Geelong,  Australia}}\\
    \IEEEauthorblockA{\textit{$^\ddagger$Centre for Smart Power and Energy Research, School of Engineering, Deakin University, Geelong, Australia}}\\
    \IEEEauthorblockA{\textit{$^\ast$Department of Computer Science \& IT, Latrobe University, Melbourne, Australia}}\\
    \IEEEauthorblockA{Email: \{mahusnoo, adnan.anwar, haftu.reda, nasser.hosseinzadeh, shama.i, robin.doss\}@deakin.edu.au, a.mahmood@latrobe.edu.au}
}

\maketitle

\begin{abstract}
 With growing security and privacy concerns in the Smart Grid domain, intrusion detection on critical energy infrastructure has become a high priority in recent years. To remedy the challenges of privacy preservation and decentralized power zones with strategic data owners, Federated Learning (FL) has contemporarily surfaced as a viable privacy-preserving alternative which enables collaborative training of attack detection models without requiring the sharing of raw data. To address some of the technical challenges associated with conventional synchronous FL, this paper proposes \textit{FeDiSa}, a novel \textit{S}emi-\textit{a}synchronous \textit{Fe}derated learning framework for power system faults and cyberattack \textit{Di}scrimination which takes into account communication latency and stragglers.  Specifically, we propose a collaborative training of deep auto-encoder by Supervisory Control and Data Acquisition sub-systems which upload their local model updates to a control centre, which then perform a semi-asynchronous model aggregation for a new global model parameters based on a buffer system and a preset cut-off time. Experiments on the proposed framework using publicly available industrial control systems datasets reveal superior attack detection accuracy whilst preserving data confidentiality and minimizing the adverse effects of communication latency and stragglers. Furthermore, we see a 35\% improvement in training time, thus validating the robustness of our proposed method.
\end{abstract}


\section{Introduction}
\label{sec:Intro}

\begin{figure*}
    \centering
    \includegraphics[width=17cm]{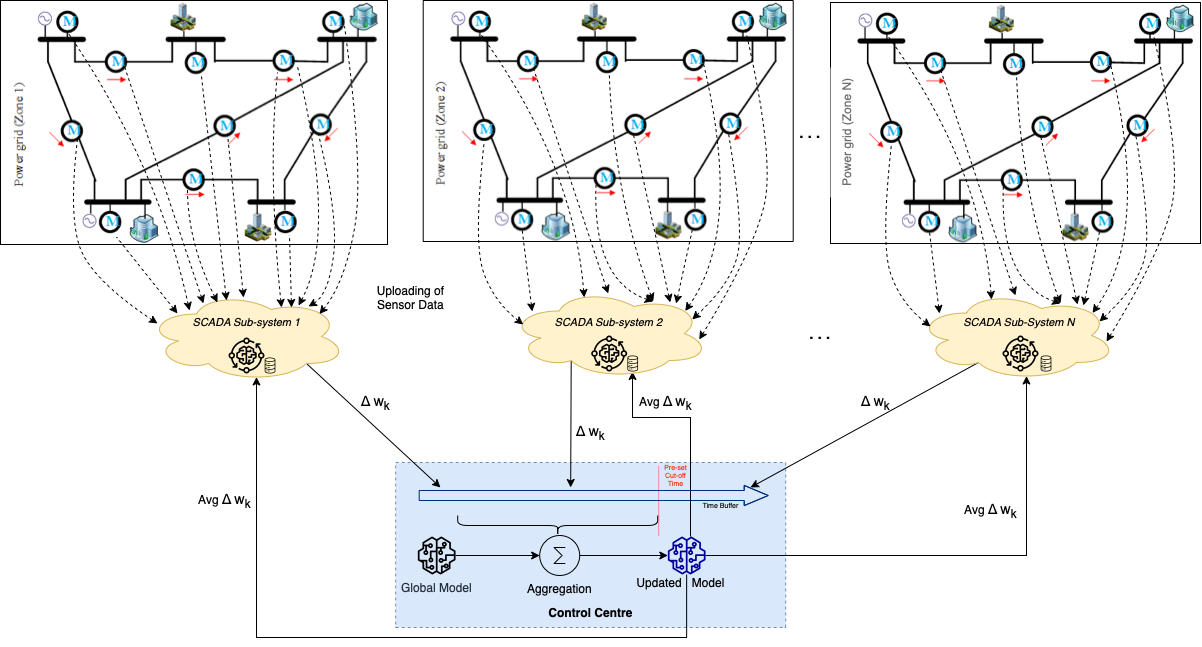}
    \caption{An overview of the FL-based intrusion detection system model. Each edge-based IED in a respective power grid zone uploads its data securely to a SCADA sub-system. Once the FL training is initiated, the control centre initializes a global model and broadcasts it to all SCADA sub-systems. Using their local datasets, the SCADA sub-systems will update the received models and send it back to the control centre. Once a preset cut-off time is reached, the control centre will start the model parameters aggregation. Any model updates received after the cut-off time will be updated into a buffer to be used for the next communication round.\label{fig:SystemFLModel}}
    
\end{figure*}

The widespread proliferation of Internet of Things (IoT) devices and groundbreaking advances in communication technology within the Smart Grid (SG) paradigm have fuelled an unprecedented increase in data generation by edge devices; however, the evolution of cyberattack crafting methods, coupled with a lack of proper attack detection strategies, has rendered the SG ecosystem an easy prey for cyberattacks launched by malicious adversaries seeking illicit financial and/or political gains \cite{Husnoo_Doss_2023}. Therefore, it is peremptory to develop effective countermeasures to protect critical assets of SGs from cyberthreats.

\subsection{Motivation}
In recent years, numerous literature \cite{Anwar_Mahmood_Ray_Mahmud_Tari_2020, 8691899, 10.1145/2806416.2806648} have proposed novel defence strategies against cyberattacks in SGs by leveraging state-of-the-art machine learning and deep learning solutions. However, such centralized cyberattack defence solutions However, such centralized cyberattack defence solutions come with limited storage capabilities, communication bottlenecks, and, most notably, privacy issues. Previous studies inherently assumed attack-free transmission of sensitive power related information which subsequently fail to provide confidentiality guarantees in real-world applications. Consequently, to address the aforementioned challenges, Federated Learning (FL)-based cyberattack countermeasures \cite{Lin_Chen_Huang_2022, 9531953, 9707860, 9540999, 10.1145/3501810, 9878267} emerged as viable privacy-preserving promising solutions which exploit the concept of  distributed learning by restricting data sharing and enabling collaborative on-device training of models at the edge. For instance, the authors in \cite{9878267} put forward a transformer-based federated false data injection attack detection mechanism and claimed superior and effective attack mitigation. 

Hitherto, prior FL-based cyberattack countermeasures make use of classic synchronous aggregation protocols (including FedAvg \cite{fedrep} and its extensions) whereby at each communication round, the central orchestrator broadcasts the model to the clients, waits for updates from all clients participating in the training and, eventually aggregates the local updates for the subsequent round until convergence. However, due to resource constraints of several Intelligent Electronic Devices (IEDs) within the real-world scenario, studies \cite{9093123} revealed few short-comings of existing FL-based cyberattack detection methodologies: 1) \textit{Stragglers}: Current state-of-the-art assume delay-free communication between the client nodes and the central orchestrator. In reality, wireless Supervisory Control and Data Acquisition (SCADA) communication transmission protocols (such as the IEC 61850) may experience delays due to unforeseen circumstances and/or unexpected dropout of participating client nodes. 2) \textit{Communication inefficiency}: Due to stragglers, the central orchestrator is required to wait for local model updates from all participating clients nodes prior to aggregation. Consequently, faster nodes are penalized due to global learning suspensions and timeouts. 3) \textit{Resource wastage}: As a result of node selection within large set-ups, multiple competent client nodes are more likely to remain idle and are unable to participate in the training process. To overcome these mentioned issues, asynchronous FL \cite{9522027, 8843942, 9606185} was proposed as an alternative to the classical synchronous FL method whereby the central orchestrator initiates the aggregation procedure without waiting for straggler updates. Nonetheless, asynchronous FL methodologies presume physical homogeneity of data on client nodes which is impractical within the SG paradigm due to perpetual data sensing by IEDs.

\subsection{Contribution and Paper Outline}
In this paper, we put forward the suitability of a semi-asynchronous federated cyberattack detection framework to discriminate between adversarial cyberattacks and natural power system disturbances within distributed power grid zone settings while considering straggler nodes. Specifically, the major contributions of this paper are summarized as follows: 1) To relax physical homogeneity and resource assumptions, we propose a light-weight semi-asynchronous privacy-preserving on-device collaborative cyberattack detection framework for power control systems, termed as \textit{FeDiSa}. 2) We leverage the use of a representation learning-based Deep Auto-Encoder model to improve the accuracy of anomaly detection in power control systems. 3) Lastly, we extensively evaluate our proposed framework on the Mississippi State University and Oak Ridge National Laboratory Power System Attack (MSU-ORNL PSA) Dataset \cite{Tommy_Morris} to validate that our proposed \textit{FeDiSa}framework still achieves high detection performance under physical heterogeneity and resource constraints. We compare our proposed framework against other state-of-the-art models and observe that our proposed solution achieves a superior detection rate (accuracy of 92.4\%) with a 35\% decrease in training time required in the presence of stragglers, which concludes the effectiveness and efficiency of our \textit{FeDiSa} framework.

The remainder of this paper is structured as follows: Section \ref{sec:probform} briefly discusses the system model and the main problem faced by synchronous FL-based cyberattack detection methods. The proposed \textit{FeDiSa} framework is detailed in Section \ref{sec:proposed}. Section \ref{sec:experiments} presents simulation scenarios for validating the efficacy of our proposed approach on publicly available datasets. Finally, Section \ref{sec:conclusion} concludes the paper.

\section{System Model and Problem Formulation}
\label{sec:probform}

\begin{figure*}
    \centering
    \includegraphics[width=17.5cm]{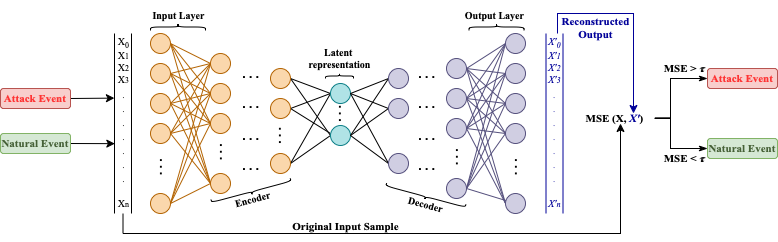}
    \caption{A schematic illustration of our attack detection module architecture with an example of a data point passing through the anomaly detection process.  The detection module consists of three components: 1) \textit{Encoder}: The encoder compresses the input into a lower dimensional latent space representation. 2) \textit{Latent Distribution}: By imposing a bottleneck in the network, we enforce a compressed knowledge representation of the original input such that sample latent vectors can be fed to the decoder. 3) \textit{Decoder}: The decoder outputs a reconstruction of the original input which is then compared against the original input. \label{fig:autoencodermodel}}
\end{figure*}

Throughout this paper, we consider a distributed power grid system which is partitioned into $N$ number of zones where $N \in \mathbb{R}^+$. As shown in Figure \ref{fig:SystemFLModel}, our proposed system model consists of three major actors: 1) \textit{Control Centre}: The control centre typically coordinates and monitors grid operations. We assume that the control centre has sufficient computational resources and acts as the central orchestrator of our federated intrusion detection framework. 2) \textit{SCADA Sub-systems}: Equipped with advanced data acquisition capabilities, SCADA sub-systems are responsible for the collection and monitoring of power data sensed by IEDs. The SCADA sub-systems will collaboratively train the federated models with their local data and update their local model parameters to the control centre. 3) \textit{Power Grid Zones}: Power grid zones consists with several IEDs and sensor networks which continually sense power-related data such as voltage phase angle, current, etc. We assume that IEDs are connected to the SCADA sub-systems via high speed communication networks for continuous transmission of sensed data.

The canonical goal of our proposed \textit{FeDiSa} framework is to solve the objective function $f_k$ for a shared global model parameter $Z \in \mathbb{R}^M$ such that $f_k(Z) = \dfrac{1}{N_k} \displaystyle \sum_{i=1}^{N_k}\ell \left(\{a_k^i, b_k^i\}; Z \right)$, where $\ell (.): \mathbb{R}^M \times \mathbb{R} \rightarrow \mathbb{R}$ is the loss to be optimized for the data point $\left(a_k^i, b_k^i\right)$. Furthermore, the global objective function $F$ over all the distributed $K$ datasets is defined as $F(Z) = \displaystyle \sum_{k=1}^K \dfrac{N_k}{N}f_k(Z)$ which is the sum of the local loss functions for each $k$, where $N$ is the total data points for training. FedAvg and its improvements \cite{10.1145/3286490.3286559} are the most common FL optimization algorithms currently in use for aggregation of local weights from all federated client nodes through multiple iterations in view of achieving model convergence.

However, as opposed to previous works on FL-based cyberattack detection \cite{9531953, Lin_Chen_Huang_2022, 9878267} which assume delay-free and error-free communication between the central FL orchestrator and the client nodes, within a real-world scenario, wireless communication protocols used for data transmission are subject to high latency which may be due to unanticipated occurrences such as weather conditions, geographical locations, high magnetic fields, etc. Indeed, for cyberattack detection within distributed grid systems, communication latency is a pressing issue that has been understudied in the context of federated learning. Additionally, active client node dropout due to power or connectivity constraints is not uncommon. The limited availability of studies that take into consideration the effect of stragglers motivates the pressing need to develop a robust and fault-tolerant FL-based framework for power system disturbances and cyberattack discrimination which is addressed by our manuscript.

\section{Proposed Method}
\label{sec:proposed}

In this section, we first present the attack detection module considered for power system disturbances and cyberattack discrimination followed by the proposed robust asynchronous FL-based intrusion detection framework. 

\subsection{Cyberattack Detection Module}

We propose to address the challenge of cyberattack detection in power network systems by leveraging the use of a Deep Auto-encoder (DAE) neural network, as schematically illustrated in Figure \ref{fig:autoencodermodel}.  DAEs are unsupervised representation learning-based models with the objective of achieving identity mapping between inputs and outputs \cite{10.1145/3097983.3098052}. Auto-encoders learn important features by parsing the input vector, $x_i$, to an encoder, $f_\alpha$, which compresses $x_j$ to a latent space representation, $y^{(j)}$, such that $y^{(j)}=f_\alpha (x_j) = a(Zx_j + c)$, where $a$ is an element-wise activation function, $Z$ is the weight matrix and $c$ is the bias vector of the encoder.  The sample latent vector, $y_j$ is then parsed to the decoder, $g_{\alpha'}$, which performs a reconstruction of the original input, $x_j$, through $x_j=g_{\alpha'}(y_i)=a(Z'y_j + c')$ while curtailing the reconstruction error. The parameters of the model $Z$, $Z'$, $c$ and $c'$ are iteratively updated during the federated training process such that $\alpha$, $\alpha' = arg \displaystyle{\min_{\alpha, \alpha'}} \ell(x_j, y_j)$ where $\ell(x_j, y_j)$ is the reconstruction error function to be optimized. 

Our DAE network architecture comprises of two symmetrical deep belief networks with five shallow layers to represent both the model's encoder and decoder. Restricted Boltzmann Machines (RBMs) are used as the building blocks of the DAE layers and are trained using Contrastive Divergence Algorithm. Rectified Linear Unit (ReLU) is used as the activation function while Mean Squared Error (MSE) is chosen as the metric for measurement of reconstruction error/anomaly degree. The anomaly threshold $\tau$ is determined by sorting the reconstruction errors for the training set in ascending order and choosing the optimal value $\tau$ at the inflection point of the error distribution. Therefore, if the reconstruction error for a certain data point exceeds $\tau$, existence of an attack event is validated. Additionally, we add soft-max layers after the RBM stack for classification. The  batch size and learning rate for training are configured to 100 and 10$^{-3}$ respectively. The model serves as the basis for cyberattack detection in the our proposed semi-asynchronous FL framework.

\subsection{Semi-asynchronous FL Framework}

As depicted in Figure \ref{fig:SystemFLModel}, throughout this manuscript, we consider a federated power system set-up whereby $K$ distributed SCADA sub-systems are linked to the control centre via communication protocols (e.g. IEC 61850, etc.). In our proposed framework, we take into consideration that the transmission protocol between the SCADA sub-systems and the control centre id not delay-free and error-free. Given the heterogeneous natural circumstances of the edge-based SCADA sub-systems, it is impossible to ensure stable communication between the control centre and the SCADA sub-systems. In such context, asynchronous FL is a viable alternative to overcome stragglers and the training time consumed by synchronous FL during the wait for all local model updates. However, such approach exacerbates the staleness of the local updates. Therefore, to remedy this problem, we propose a semi-asynchronous framework that relies upon a preset cut-off time for model updates aggregation. 

\begin{algorithm}
\textbf{Input}: learning rate $\eta$, one-bit quantizer $sign(.)$, $K$ SCADA sub-systems, Local training set $D_k$ for $k \in [1, K]$
\vskip0.5em
\hskip0.5em \textit{Initialize} common unanimous model parameter $Z_0$, model parameter buffer $B$, aggregation time $T_a$, SCADA sub-systems time cost [$s_0, s_1, ..., s_k$].
\bigbreak

\hskip0.5em \For{each communication round $t$ in $T_{cl} \in (1,n)$}{
\bigbreak
\textbf{Control Centre}:

\hskip0.5em \textit{Obtain} gradient ($g^t_k$, $t_k$) from each SCADA sub- system $k$.
\vskip0.5em
\hskip0.5em $s_k \leftarrow$ update $t_k$.
\vskip0.5em
\hskip0.5em $B \xleftarrow{+}$ ($g^t_k$, $t_k$)
\vskip0.5em
\hskip0.5em \If {Aggregation time $T_a$ is reached}{
\vskip0.5em
 $B' \leftarrow$ cluster by model version $t$ and compute aggregation. 
\vskip0.5em
\hskip0.5em \For{each ($g^t_k$, $t_k$) in $B'$}
{
\vskip0.5em
\textit{Update} new global $g$ and $t$.
\vskip0.5em
\textbf{end}
}
\vskip0.5em
\textit{Update} newly aggregated global parameters ($g$, $t$) to source SCADA sub-systems $k$.
\vskip0.5em
\textit{Clear} $B$.
\vskip0.5em
\textbf{end}
}


\bigbreak

\textbf{SCADA Sub-System}:

\hskip0.5em \For{each $k \in [K]$}{
\vskip0.5em
\textit{Compute} local gradient $g^t_k = \nabla f_k(Z^t)$ using $D_k$. 
\vskip0.5em
\textit{Compute} time cost of each client $t_k$
\vskip0.5em
\textit{Update} ($g^t_k$, $t_k$) to control centre. 
\vskip0.5em
\textbf{end}
}
}
\textbf{end}
\vskip0.5em
\textbf{Output}: Global Model

\caption{Proposed \textit{FeDiSa} Framework}
\label{algo:federated}
\end{algorithm}

During a particular $t$th federated training round, our proposed \textit{FeDiSa} framework, as presented in Algorithm \ref{algo:federated} consists of the following implementation phases: 1) The control centre distributes the global DAE model parameters $Z^t$ to all $K$ SCADA-subsystems. 2) Each SCADA sub-system $k$ for $k \in$ [$k$] then computes the local model gradient $g^t_k = \nabla f_k(Z^t)$ utilizing its local training set. 3) Each $k$ SCADA sub-system updates its locally computed gradient $g^t_k$ to the control centre. 4) The control centre updates the model parameter buffer and the time cost for each $k$ SCADA sub-system. 5) Once the preset cut-off time $T_a$ is reached, the control centre clusters the model updates by the time version and aggregates the updates of each group using the method proposed in \cite{https://doi.org/10.48550/arxiv.1903.03934}. 6) Lastly, the control centre then updates the new model parameters to each SCADA-subsystem $k \in [K]$. 

\section{Experimental Results and Discussions}
\label{sec:experiments}

In this section, we first discuss the dataset used for the experimental set-up followed by the achieved results that validate the attack detection performance and the robustness of our proposed \textit{FeDiSa} framework detailed in Section \ref{sec:proposed}.

\subsection{Dataset Description and Preparation}

We use the publicly available industrial control system MSU-ORNL PSA dataset \cite{Tommy_Morris} to experimentally evaluate the effectiveness of our proposed \textit{FeDiSa} framework.  The datasets are a binary class comprised of 15 distinct data files, each with a different power system event scenario. The event scenarios include natural events (power system disturbances) and attack events (cyberattacks). The natural events include faults of varying level across the distributed grids, namely 10-19\%, 20-79\%, and 80-90\%. Similarly, the attack events are as follows: 1) A remote tripping command injection (i.e., an attack vector that involves sending a command to an IED which renders a circuit breaker to open/close). 2) A change in IED settings  (i.e., an attack vector targeting the IEDs to change their setting to disable their normal operation, for example, to falsely trip a relay). 3) Then, a false data injection attack (i.e., an attack vector that intelligently emulates a valid physical fault by modifying power system values across voltage, current, and other physical parameters to blind system operators and render consequential impacts). The datasets have been simulated and recorded based on 128 features (such as frequency, current phase angle, current phase magnitude, etc.).  

A basis of 100 features have been chosen for training our DAE-based attack detection model after applying Principal Component Analysis (PCA) to cull out the unnecessary features and reduce training complexity. To deal with missing values within the dataset, we use K-Nearest Neighbour (KNN) imputation. The input features are normalized in the range of (0,1). Lastly, we split the dataset into a train-test split ratio of 70\% and 30\% respectively.

\begin{table}[!h]
\caption{Comparison of the average attack detection performance of our proposed approach against state-of-the-art models. \label{tablecompa}}

\begin{tabular}{l|lllll|}
\cline{2-6}
& \multicolumn{5}{c|}{\textbf{Federated Models}}        \\ \hline
\multicolumn{1}{|l|}{\textbf{Metric (\%)}}              & \multicolumn{1}{l|}{\textit{\textbf{Prop. Approach}}} & \multicolumn{1}{l|}{\textit{\textbf{CNN}}} & \multicolumn{1}{l|}{\textit{\textbf{LSTM}}} & \multicolumn{1}{l|}{\textit{\textbf{RBM}}} & \textit{\textbf{RNN}} \\ \hline
\multicolumn{1}{|l|}{\textit{\textbf{Accuracy}}}   & \multicolumn{1}{l|}{92.4}                                & \multicolumn{1}{l|}{89.6}                  & \multicolumn{1}{l|}{78.5}                   & \multicolumn{1}{l|}{73.3}                  & 69.3                  \\ \hline
\multicolumn{1}{|l|}{\textit{\textbf{Precision}}}  & \multicolumn{1}{l|}{90.5}                                & \multicolumn{1}{l|}{87.7}                  & \multicolumn{1}{l|}{79.1}                   & \multicolumn{1}{l|}{72.5}                  & 68.8                  \\ \hline
\multicolumn{1}{|l|}{\textit{\textbf{Recall}}}     & \multicolumn{1}{l|}{87.9}                                & \multicolumn{1}{l|}{85.9}                  & \multicolumn{1}{l|}{78.3}                   & \multicolumn{1}{l|}{71.7}                  & 69.5                  \\ \hline
\multicolumn{1}{|l|}{\textit{\textbf{F1-Measure}}} & \multicolumn{1}{l|}{88.4}                                & \multicolumn{1}{l|}{87.6}                  & \multicolumn{1}{l|}{79.2}                   & \multicolumn{1}{l|}{71.3}                  & 69.7                  \\ \hline
\end{tabular}

\end{table}

\subsection{Attack Detection Performance}
\begin{figure}
    \centering
    \includegraphics[width=6cm]{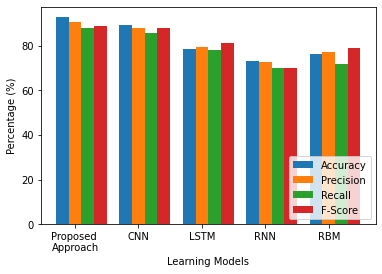}
    \caption{Comparison of the average attack detection performance of our proposed framework against four competing deep learning models based on percentage accuracy, precision, recall and F-Score metrics.\label{fig:metriccompaall}}
\end{figure}

\begin{figure}
    \centering
    \includegraphics[width=6cm]{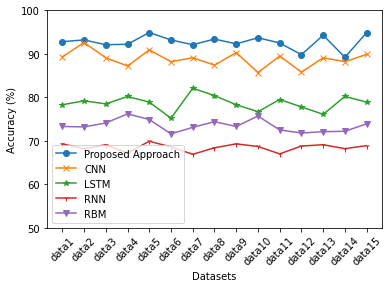}
    \caption{Comparison of the average attack detection accuracy of our proposed approach against four other neural networks (trained in similar fashion) over different datasets.\label{fig:accccompawholedata}}
\end{figure}

We validated the attack detection performance of our proposed semi-asynchronous FL-based cyberattack detection framework against four other state-of-the-art models used for intrusion detection in SGs namely Convolutional Neural Network (CNN), Long Short-Term Memory (LSTM), Recurrent Neural Network (RNN) and Restricted Boltzmann Machine (RBM). The rivalling models are all trained in a similar semi-asynchronous FL-based configuration. After training, the global models are evaluated on the test set.  From 
Figure \ref{fig:metriccompaall} and Table \ref{tablecompa}, we evaluate the average attack detection performance of our proposed approach using four different classification metrics namely accuracy, precision, recall and F1-Measure. The experimental validations reveals that our proposed \textit{FeDiSa} framework achieves superior average cyberattack detection accuracy over other state-of-the-art deep learning models. For instance, our proposed approach achieved an accuracy rate of 92.4\% in comparison with CNN (89.6\%), LSTM (78.5\%), RBM (73.3\%) and RNN (69.3\%). Similarly, in terms of precision rate, our proposed approach achieved 90.5\% as opposed to CNN (87.7\%), LSTM (79.1\%), RBM (72.5\%) and RNN (68.8\%). In addition, we observe that RNN produces the lowest detection performance overall. Next, we evaluate the cyberattack detection rate of our proposed framework using the fifteen different data files within the MSU-ORNL PSA dataset. From the experimental results in Figure \ref{fig:accccompawholedata}, we observe that our proposed framework outperforms other neural networks over all 15 data files. For example, the accuracy of FeDiSa on data1 is 92.6\% as compared to CNN (88.9\%) or LSTM (78.4\%). Similarly, for data15, we achieve improved accuracy detection rate of 94.9\% as opposed to RNN (68.9\%) and RBM (73.9\%). Therefore, we conclude that our proposed semi-asynchronous FL-based power system disturbance and cyberattack discrimination framework achieves sufficient attack detection accuracy as opposed to other deep learning models whilst guaranteeing data confidentiality and minimizing the effect of communication delays and stragglers.

\begin{figure}
    \centering
    \includegraphics[width=6cm]{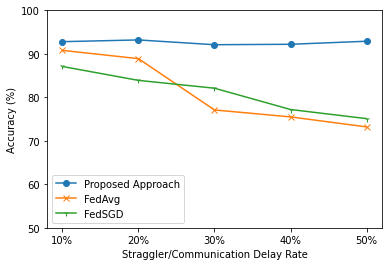}
    \caption{Comparison of the robustness of our proposed semi-asynchronous approach against two classical synchronous FL aggregation methods based on accuracy with increasing affected nodes. \label{fig:comparobust}}
\end{figure}

\subsection{Robustness against Stragglers}

Taking into account the effect of communication latency and straggler client nodes in a real-world Smart Grid scenario, we propose a semi-asynchronous FL-based approach to train a global cyberattack detection model. We set the cut-off aggregation time to twice the average waiting time of nodes in a federated set-up. To simulate the effect of stragglers and communication latency, we initially set $K$ to 10 and set a pause mechanism for a set portion of SCADA sub-systems such that they will be non-responsive for that period of time. We compare the performance of our proposed approach against that of two classical synchronous FL aggregation solutions, FedAvg and Federated Stochastic Gradient Descent (FedSGD). From Figure \ref{fig:comparobust}, we observe that the accuracy rate of cyberattack detection of our proposed approach stays relatively consistant with increasing number of affected nodes. On the other hand, the synchronous FL aggregation protocols oversee a significant decline in their performance with increasing number of impacted SCADA sub-systems. This signifies that our semi-asynchronous attack detection solution is robust to communication lags and straggler SCADA sub-systems. 

\begin{figure}
    \centering
    \includegraphics[width=6cm]{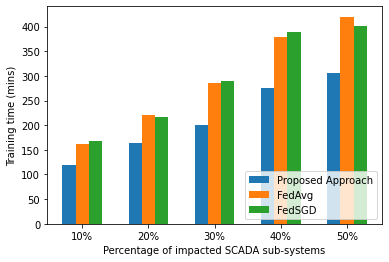}
    \caption{Comparison of training time of our proposed approach against FedAvg and FedSGD on varying number of impacted SCADA sub-systems. \label{fig:camptime}}
\end{figure}

Furthermore, we evaluate the training time taken to achieve model convergence within a classical Federated Learning environment using FedAvg and FedSGD in different scenarios of affected nodes. As shown in Figure \ref{fig:camptime}, we note that the training time of our proposed method in the presence of stragglers and communication delays is reduced as compared to the synchronous FedAvg and FedSGD aggregation algorithms. Specifically, it is highlighted that the training time for our proposed \textit{FeDiSa} solution is approximately 35\% less that the rivalling aggregation methods. However, we also notice that there is an increase in the training time of our proposed solution with increasing number of impacted SCADA sub-systems. This is be due to the increasing number of training rounds required to achieve model convergence with an increase in affected SCADA sub-systems nodes.  

\section{Conclusion}
\label{sec:conclusion}

This paper proposes a novel semi-asynchronous FL-based power system detection and cyberattack detection strategy to mitigate the effects of communication unreliability and stragglers in real-world SG applications on decentralized power grid systems. Furthermore, a representation learning-based DAE model is leveraged for timely and accurate detection of cyberthreats. Experimental validations performed on a publicly available dataset (the MSU-ORNL PSA dataset) revealed that the proposed solution outperforms other deep learning models in terms of detection accuracy. while being computationally efficient and achieving a  35\% reduction in training time compared to state-of-the-art synchronous solutions, thus proving the effectiveness and efficiency of the proposed \textit{FeDiSa}. For future works, we will explore security aspects of FL-based detection frameworks, which are highly vulnerable to byzantine threats as per our previous work \cite{https://doi.org/10.48550/arxiv.2209.14547}.

\bibliographystyle{IEEEtran}
\bibliography{refs.bib}

\end{document}